\documentstyle[preprint,aps]{revtex}
\begin{document}
\draft
\title{Shakeup spectrum in a two-dimensional electron gas in a strong magnetic field.}
\draft
\author{E. Tsitsishvili, Y. Levinson}
\address{Department of Condensed Matter Physics, The Weizmann Institute of Science,\\
 76100 Rehovot, Israel}
\draft \maketitle

\begin{abstract}

The shakeup emission spectrum in a two-dimensional electron gas in a strong magnetic field is
calculated analytically. The case of a localized photocreated hole is studied and the
calculations are performed with a Nozi\` eres-De Dominicis-like Hamiltonian.
The hole potential is assumed to be small compared to the cyclotron energy and is therefore
treated as a perturbation.
Two competing many-body effects, the shakeup of the electron gas in the optical transition, and
the excitonic effect, contribute to the shakeup satellite intensities.
It is shown, that the range of the hole potential essentially influences the shakeup spectrum.
For a short range interaction the above mentioned competition is more important and
results in the shakeup emission quenching when electrons occupy only the lowest Landau level.
When more than one Landau level is filled, the intensities of the shakeup satellites
change with magnetic field nonmonotonically.
If the interaction is long range, the Fermi sea shakeup processes dominate.
Then, the satellite intensities
smoothly decrease when the magnetic field increases and there is no suppression
of the shakeup spectrum when the only lowest Landau level is filled.
It is shown also that a strong hole localization is not a necessary condition for the SU spectrum
to be observed. If the hole localization length is not small compared to the magnetic length,
the SU spectrum still exists. Only the number of contributions to the SU spectrum
reduces and the shakeup processes are always dominant.\\
$\dag$Permanent address: Institute of Cybernetics,Academy of Sciences of Georgia,Euli 5, 380086\\
Tbilisi,Georgia
\end{abstract}
\maketitle
\newpage
\section{I\lowercase{ntroduction}}
\label{sec:in}
It is known that many body processes can have a dramatic effect on the optical spectra
of a degenerate electron gas. One of the most notable examples is optical absorption and
emission processes  in metals where a localized hole state is involved, the so-called X-ray
Fermi edge singularity (FES) phenomena \cite{M1}-\cite{G}.
This effect is dominated by two competing contributions, the Anderson
orthogonality catastrophe \cite{A} due to the Fermi sea shakeup and the Mahan
``exciton'' \cite{M2}, due to the electron-hole interaction. The first effect
leads to a strong reduction of the absorption and the emission intensity close to the X-ray
absorption edge, while the second one is manifested by the power law divergence at the
threshold energy \cite{ND}. The shakeup effect that occurs in optical transitions
in which excitations of the electron gas are created, also leads to a low-energy
tail in the emission spectrum \cite{CO}.
The same type of processes were studied in semiconductor quantum well structures
with a degenerate two-dimensional electron gas (2DEG). The FES
effect in the luminescence spectra of modulation-doped InGaAs-InP quantum wells was
observed in Refs. \cite{Sh1}, \cite{N}. The effect was attributed to the hole
localization due to the alloy fluctuations.

The experiments in 2DEG structures were also carried out in the quantum Hall regime.
The case of a 2DEG in a semiconductor in a strong perpendicular magnetic field
essentially differs from that of metals, since in the 2D case the conduction-band states are
quantized into discrete Landau levels (LLs). As a result the photoluminescence spectrum turns
into a series of equally spaced magnetoluminescence (ML) lines with the spacing
given by the electron cyclotron energy $\hbar \omega_{B} = e \hbar B/m_{e} c$.
In addition, the low-energy photoluminescence tail is also transformed into a series of
satellite lines, so-called shakeup (SU) satellites, which are below the main ML spectrum.
The reason is that in the presence of a magnetic field the shakeup process generates discrete
inter-LL excitations, ``magnetoplasmons''
(MPs), in which an electron is promoted from one LL to a higher one, across the Fermi energy.

For the first time, the FES effect (a strong enhancement of the intensities of the ML
spectral lines towards the Fermi energy) and the SU satellites were reported in
Ref. \cite{Sh1} and Ref. \cite{B}, respectively. Futher experimental investigations of
the SU spectra in 2D structures were carried out in Refs.\cite{NS}-\cite{Os}.
In early Refs.\cite{N},\cite{NS} was found that the strength of the shakeup is controlled by the
localization of the photocreated hole. In addition,  a strong suppression of shakeup
processes at hight magnetic fields was observed and qualitatively discussed
in Refs.\cite{N},\cite{NS} . The discussion is based
on the theoretical results for the hole Green's function at zero magnetic field which are
obtained in Ref.\cite{Lan}. In recent Refs.\cite{Me},\cite{Me1} periodic changes of the shakeup line
intensities in the filling factor were reported. In Ref.\cite{Me2} a role of the hole localization on
the main and SU magnetoluminescence spectra was studied.
In Ref.\cite{BJ}  2D structures with delocalized photocreated holes were investigated and
a strong suppression of shakeup processes was observed when electrons occupy only the lowest LL.
The shakeup processes due
to the creation of the spin-wave excitations were studied in Refs.\cite{Pl},\cite{Osb}.

A theory of optical and magneto-optical phenomena in a 2DEG was given in Refs. \cite{Hw}-
 \cite{Tav}, where numerical calculations of the optical spectra are presented.
In Ref.\cite{US} the influence of the hole mass, temperature and the electron-hole interaction
strength on the main magneto-optical spectra is discussed. It is noted also that
the SU spectrum is more pronounced when the interaction strength increases.
Recently the FES problem in a 2DEG was solved analytically for the case of a weak magnetic
field when a large integral number of Landau levels are filled \cite{W}.
In Ref. \cite{BJ} the theoretical model
of the SU spectrum was developed for the case of an unlocalized photocreated hole.
Observations of a strong suppression of the shakeup effects at a
filling factor close to $\nu = 1$ are  explained by the competition between the interband and
intraband scattering processes.

An analytical description of the SU spectrum for the case of a localized hole is
absent to-day, to our knowledge. At that reason, some theoretical problems are open up to now.
For example, the role of the excitonic effect and hole localization on the SU spectrum is
not clear. Usually, it is assumed that the SU spectrum is due to the Anderson
orthogonality catastrophe and the contribution of the Mahan effect is ignored
(see, e.g., Refs. \cite{N}, \cite{NS}). On the other hand, in Ref.\cite{W} is shown that
in the case of
the main absorption and emission spectra a balance between the Mahan effect and the
orthogonality catastrophe is important, similarly to the case of the X-ray problem in metals.
In the early experimental works \cite{N}, \cite{NS} was supposed also that both the FES effect
and SU spectrum are very sensitive to the localization degree of the photocreated hole, as noted
above. However, this assumption is in contrudiction to the numerical calculations in
Ref.\cite{US} and the experimental results in Refs.\cite{BJ},\cite{Me2}.
In addition, in experiments some pecularities of the SU spectrum noted above
(a strong suppression of shakeup processes at high magnetic fields\cite{N},\cite{NS} and
the periodic changes of the shakeup line intensities in the filling factor\cite{Me},\cite{Me1})
are observed. These properties of the SU spectrum can not be explain by a simple discussion given in
Refs.\cite{N}, \cite{NS} and require an analytical treatment.

In this paper we present an analytical study of the SU emission spectrum
due to transitions between a localized
level and a 2DEG, subjected to a strong magnetic field. Calculations of the shakeup
spectrum are performed with a Nozi\` eres - De Dominicis-like Hamiltonian \cite{Hw}, \cite{W},
in which we treat the interaction part as a perturbation.
The electron-hole interaction is assumed to be much smaller than the cyclotron energy. In this
case the SU spectrum can be calculated perturbatively, as the SU satellites are due to the creation
(by the optical transition) of the electron-"hole" pairs at different Landau levels.
We show that the SU spectrum is determined by both the Anderson orthogonality and the Mahan effect.
In the general case the contribution due to the Fermi sea
shakeup is dominant. However, for the special case of a short
range hole potential, the excitonic effect competes with the shakeup effect. This competition
results to a strong supression of the SU spectrum if electrons occupy only the lowest Landau level.
We show also that a strong hole localization is not a necessary condition for the SU spectrum to be
observed. In addition, we discuss the filling factor
dependence of the SU lines, as well as the influence of the hole potential range.

\section{Problem formulation}
\label{sec: pf}
We consider here a model that is similar to the one used in the study of core level optical
spectra in metals \cite {M2}, \cite{ND}, \cite{SS}, since most of the
experiments on magneto-optical spectra in a 2DEG were carried out in samples
with a disorder due to composition fluctuations. In such
structures some amount of holes are localized due to alloy fluctuations and behave
like the core holes in the X-ray problem.  Note that a very small hole localization length
of $10 {\rm \AA} - 30 {\rm \AA}$ was reported in Ref.\cite{NS} for an InGaAs/InP quantum well.
For comparison,  in GaAs crystals the magnetic length  $l_{B} \geq 50 {\rm \AA}$ at
a magnetic field of $B \leq 25T$.

Thus, we assume that the photocreated hole state is nondegenerate, and strongly localized, with a
localization length $a$ much smaller than the magnetic length $l_{B}$.
The electron-electron interaction
in the conduction band is ignored and the Hamiltonian of the problem consists of three
contributions \cite{Hw}, \cite{W}, \cite{SS}.

\begin{eqnarray}
H = H_{0} + H_{h} + H_{int}.
\label{H}
\end{eqnarray}
The first term describes noninteracting electrons, taken here as spinless, and the operator
$c_{n}^{\dag}$ creates an electron
in the Landau state $|n \rangle$ with an energy of $\varepsilon_{n}$,

\begin{eqnarray}
H_{0} =  \sum_{n} \varepsilon_{n} c_{n}^{\dag} c_{n}.
\label{H0}
\end{eqnarray}
The second term describes the hole, and the operator $d^{\dag}$ creates a hole in a
nondegenerate state $|h\rangle$, with a negative energy $E_{0}$,  measured from the bottom of
the conduction band at $B=0$,

\begin{eqnarray}
H_{h} = E_{0} d^{\dag} d.
\label{Hh}
\end{eqnarray}
The third term describes the coupling between the conduction
electrons and the hole which creates a potential  $\; V(\vec{r})$,

\begin{eqnarray}
H_{int} =  d^{\dag} d \; V, \qquad V = \sum_{n, n'} V_{n n'}  c_{n}^{\dag} c_{n'}.
\label{Hi}
\end{eqnarray}

The emission spectrum is given by the real part of the Fourier transform of the response function

\begin{eqnarray}
I(\omega) = {\cal{N}}\times {\cal{R}}e \;\int_{0}^{\infty} dt \; e^{-i\omega t}\;
{\cal{F}}^{e}(t), \qquad
{\cal{F}}^{e}(t - t') = \sum_{n n'} M_{n} M_{n'}^{\star}\; F_{n n'}^{e}(t - t'),
\label{S}
\end{eqnarray}
where $ {\cal{N}}$ is a normalization constant.
The optical matrix element $M_{n}$ describes the electron-photon coupling,
$M_{n} = P_{vc} \; \langle n|h \rangle$, where the interband momentum  matrix element $P_{vc}$
incorporates the effect of the lattice periodicity and $ \langle n|h \rangle$ is the overlap
between the electron and hole envelope functions. The two-particle Green's function is given by

\begin{eqnarray}
F_{n n'}^{e}(t - t') =  \langle i_{h}^{e} | T \{ c_{n}^{\dag}(t)
 d^{\dag}(t) d(t') c_{n'}(t')\} | i_{h}^{e} \rangle,
\label{F}
\end{eqnarray}
where $T$ is time-ordering operator,  the operators $c_{n}(t) = e^{i H t} c_{n}(0) e^{-i H t}$,
and $d(t) = e^{i H t} d(0) e^{-i H t}$, are taken in the Heisenberg representation.
The state $| i_{h}^{e} \rangle$ is the initial state of the electron system in the presence of a
localized hole. This state  is a product of the hole state
$| h\rangle$, and the Slater determinant $| i^{e} \rangle$ of
single-particle electron states, which differ from the Landau electron states $| n\rangle$
due to the hole potential, and are the eigenstates of the Hamiltonian $H^{e} = H_{0} + V$.

It was shown in Ref. \cite {ND} that the calculation of the optical  spectra reduces
to a  `` one-body''
problem when the conduction electrons are scattered by a ${\it{transient}}$ potential due
to the hole,  switched on/off suddenly at the time of the transition.
As a result the two-particle Green's function in the time representation
is ${\it {exactly}}$ a product of
the hole Green's function, which accounts for the shakeup effect, and
the so-called transient electron Green's function which accounts for the exciton effect,

\begin{eqnarray}
F_{n n'}^{e}(t - t') = - g_{n' n}^{e}(t', t|t, t')\; D^{e}(t - t').
\label{FL}
\end{eqnarray}
In Eq. (\ref{FL}) the hole Green's function is but the average of the evolution
operator,

\begin{eqnarray}
D^{e}(t-t') =  e^{- i E_{0} (t - t')} \;\langle i^{e} | \hat{S}(t, t') | i^{e}  \rangle, \qquad
 \hat{S}(t, t') =  T \; \exp\{i \int_{t'}^{t} d \tau \tilde{V}(\tau) \},
\label{Ss}
\end{eqnarray}
where  $\tilde{V}(\tau)$  is the interaction $V$ from Eq. (\ref{Hi}),
in the Heisenberg representation with the Hamiltonian $H^{e}$.
The  transient electron Green's function is given by

\begin{eqnarray}
g_{n n'}^{e}(\tau, \tau^{'}|t, t')  =
\frac{\langle i^{e} | T \{ \tilde{c}_{n}(\tau) \tilde{c}_{n'}^{\dag}(\tau^{'})\;
\hat{S}(t,t') \}| i^{e} \rangle}
{\langle i^{e} |\hat{S}(t,t')| i^{e} \rangle},
\label{tf}
\end{eqnarray}
and it satisfies the Dyson equation, that is known in the X-ray problem as the Nozi\` eres -
De Dominicis equation \cite{ND}:

\begin{eqnarray}
g_{n n'}^{e}(\tau, \tau^{'}|t, t') = G_{n n'}^{e}(\tau - \tau^{'}) +
i \; \int_{t'}^{t} d \tau^{''} \sum_{s s'}\; G_{n s}^{e}(\tau - \tau^{''})\;
V_{s s'} \; g_{s' n'}^{e}(\tau^{''}, \tau^{'}|t, t').
\label{ND}
\end{eqnarray}
The conduction electron Green's function $G_{n n'}^{e}(t)$ entering Eq. (\ref{ND}) is defined by
\begin{eqnarray}
G_{n n'}^{e}(t - t') =
\langle i^{e} |T\{\tilde{c}_{n}(t)\; \tilde{c}_{n'}^{\dag}(t')\}| i^{e} \rangle,
\label{G}
\end{eqnarray}
and it obeys the Dyson equation,

\begin{eqnarray}
G_{n n'}^{e}(t) = G_{n}^{(0)}(t)\;\delta_{n n'} -
i \; \sum_{n''}\;\int_{-\infty}^{+\infty} d \tau \; G_{n}^{(0)}(\tau)\;
V_{n n''}\; G_{n'' n'}^{e}(t - \tau),
\label{GD}
\end{eqnarray}
where the bare electron propagator

\begin{eqnarray}
G_{n}^{(0)}(t) = e^{-i\varepsilon_{n}t}\;\Bigl[\theta(t) -
n(\varepsilon_{n})\Bigr],
\label{G0}
\end{eqnarray}
and $n(\epsilon_{n})$ are the electron occupation numbers.

\section{General results}
\label{sec: ptr}

As was noted above, the low-energy satellites of the emission spectrum
are supposedly due to the shakeup effect, when the optical transition creates
 ``magnetoplasmons'' \cite{N},\cite{NS}.
(In what follows the MPs are the free electron-``hole'' pairs, since
 the electron-electron interaction in the conduction band is neglected.)
The energies which are needed to
create the MPs  are determined by the cyclotron energy $\hbar \omega_{B}$.
Therefore, one can expect that in the case of a strong magnetic field the interaction
 Hamiltonian $H_{int}$ in Eq. (\ref{H}) can be treated as a perturbation if the hole potential
is much smaller than $\hbar \omega_{B}$. Following this approximation, we
calculate the hole Green's function (\ref{Ss}) and the transient electron Green's
function (\ref{tf}) up to the second-order in the interaction potential $H_{int}$, Eq. (\ref{Hi}),
by iterations of Eqs. (\ref{GD}) and (\ref{ND}), respectively. Then, we collect the
same order terms in the two-particle Green's function (\ref{FL}) and perform its Fourier
transform. The hole level shift due to the electron-hole interaction is ignored.
Finally, we find that the corrections to the main ML spectrum appear already in the
first order,

\begin{eqnarray}
I_{ML}(\omega) = {\cal{N}} \; \sum_{n} \delta(\hbar \omega - \hbar \omega_{n})\;
n(\varepsilon_{n})\;
\Bigl\{|M_{n}|^2 + 2\;M_{n}\;\sum_{s}
\frac{V_{n s}}{\varepsilon_{n} - \varepsilon_{s}}\;M_{s}^{\star}\;
[1 - n(\varepsilon_{s})]\Bigr\}.
\label{PE}
\end{eqnarray}
The first term in  Eq. (\ref{PE}) is a well known result for the emission spectrum in the
one-electron description of the optical process. All the occupied LLs with $M_{n} \neq 0$
contribute to this term and the energy $\hbar \omega_{n} = |E_{0}| + \varepsilon_{n}$
determines the spectral position of the $n$th ML line.
The second term is due to the lowest-order many body corrections to the main ML spectrum.
This term has its origin in the transient electron Green's function and determines the change of
the ML line intensity due to the hole potential. At zero temperature quantum numbers $n$ and
$s$ refer to different LLs, $n$ refers to occupied states, while $s$ to empty ones.
The closer the $n$th ML line is to the Fermi energy, the larger is its intensity change,
due to the corresponding energy differences in the denominators.

Unlike the main spectrum Eq. (\ref{PE}), the shakeup emission spectrum is absent in a
single-particle approximation,  since it is due to many-body effects. We
find that the SU spectrum appears in the second order of the perturbation and is given by

\begin{eqnarray}
I_{SU}(\omega) = {\cal{N}} \;
\sum_{n,n',s} \delta\Bigl(\hbar \omega - \hbar \omega_{n} +
(\varepsilon_{s} -\varepsilon_{n'})\Bigr)\;
\Bigl\{|M_{n}|^2\;\frac{|V_{s n'}|^{2}}{(\varepsilon_{s} -
\varepsilon_{n'})^{2}}\;  \nonumber\\
- M_{n}\;\frac{V_{n s} V_{s n'}}{(\varepsilon_{s} - \varepsilon_{n})
(\varepsilon_{s} - \varepsilon_{n'})}\;M_{n'}^{\star}\Bigr\} n(\varepsilon_{n})\;n(\varepsilon_{n'})\;[1 - n(\varepsilon_{s})].
\label{SU}
\end{eqnarray}
As in Eq. (\ref{PE}), for zero temperature the  quantum numbers $s$ refer to empty
states, while $n$ and  $n'$ refer to
occupied ones. Therefore, the $\delta-$functions in  Eq. (\ref{SU}) ensure that the
shakeup satellites appear below the main ML peaks. The first term in the brackets
arises from the hole Green's function (\ref{Ss}),  and describes
the  Fermi sea shakeup effect in the emission process.
The second term  has its origin in the electron transient propagator
(\ref{tf}),  and is due to the interaction of the hole with the extra electron that is created
in the optically active Landau state with $M_{n}\neq 0$,  in the process of absorption.
Thus,  the SU spectrum is a result of a balance between the contribution of the
orthogonality catastrophe and that of the excitonic effect.

In what follows we present a more detailed study of the SU spectrum,  Eq. (\ref{SU}).
To proceed, we rewrite this equation  using the axial gauge
representation for the Landau states $|n\rangle = |Nm\rangle$
with a center at the hole position, where $N$ is the Landau number and $m$ is
the angular momentum. Then we arrive at

\begin{eqnarray}
I_{SU}(\omega) = {\cal{N}} \; \sum_{k=1}^{\infty}\;
\delta\Bigl(\hbar \omega - \hbar \omega_{0} + k \hbar \omega_{B}\Bigr)\;
\Bigl[{\cal {A}}_{k} - {\cal {M}}_{k}\Bigr],
\label{SUm}
\end{eqnarray}
where

\begin{eqnarray}
{\cal {A}}_{k} = \sum_{N,m} \sum_{N', m'} \sum_{m_{s}}\;
|M_{Nm}|^2\;\frac{|V_{(N+N'+k)m_{s}, N'm'}|^{2}}{(\varepsilon_{N+N'+k} -
 \varepsilon_{N'})^{2}}\; F_{NN'k},  \nonumber\\
{\cal {M}}_{k} = \sum_{N,m} \sum_{N', m'} \sum_{m_{s}}\;
 M_{Nm}\;\frac{V_{Nm, (N+N'+k)m_{s}} V_{(N+N'+k)m_{s}, N'm'}}
{(\varepsilon_{N+N'+k} - \varepsilon_{N})
(\varepsilon_{N+N'+k} - \varepsilon_{N'})}\;M_{N'm'}^{\star}\; F_{NN'k},
\label{SAM}
\end{eqnarray}
with

\begin{eqnarray}
F_{NN'k} = n(\varepsilon_{N})\;n(\varepsilon_{N'})\;[1 - n(\varepsilon_{N+N'+k})].
\label{SAM1}
\end{eqnarray}

It is clearly seen from Eq. (\ref{SUm}) that the $k$th satellite $(\rm{SU}_{k})$ of
the shakeup spectrum  appears at an energy  $ k \hbar \omega_{B}$ below the lowest-energy
main  ML line of energy $\hbar \omega_{0} = |E_{0}| + \frac{1}{2} \hbar \omega_{B}$.
In Eq. (\ref{SAM}) the summation over the LL numbers  $N$ and  $N'$
refers to the occupied LLs. In the shakeup  term ${\cal {A}}_{k}$ the
number $N$ indicates the LL which contributes to
the emission process, while  $N'$ indicates the LL which is involved in the
MP creation. Note that the state $|N'm'\rangle$ is not necessarily optically
active. In the exciton term ${\cal {M}}_{k}$, however, the electron state that is involved in
the MP creation is always optically active. Therefore, evidently, all channels contribute to
the MP creations, except those which contribute to the optical transitions.
Indeed, from
Eq. (\ref{SAM}) it follows that the terms with $N=N'$ and $m=m'$ in ${\cal {A}}_{k}$ and
${\cal {M}}_{k}$ cancel each other. The remaining terms in ${\cal {M}}_{k}$
are nonzero if the LLs with $N \neq N'$ are optically active, i.e., $M_{N} \neq 0$ and $M_{N'} \neq 0$.

As it follows from Eq. (\ref{SUm}),  each SU satellite results from a superposition
of replicas from different ML lines.
As an illustration,  the possible options for the first ${\rm{SU}_{1}}$ satellite are shown
schematically in Fig.~\ref{fig.1}. Electrons occupy the three lowest LLs which are assumed to
be  optically active. Thus, the ML spectrum,  Eq. (\ref{PE}),  contains three lines, their
corresponding transitions are shown by the
solid arrows in Fig.~\ref{fig.1}. Each of these transitions contributes to the ${\rm{SU}_{1}}$
satellite at $\hbar \omega_{0} - \hbar \omega_{B}$, if it is accompanied by a MP excitation, shown
by the dashed arrows. The ML line at $\hbar \omega_{0}$ contributes to the ${\rm{SU}_{1}}$
satellite if it is accompanied by an $\hbar \omega_{B}$ MP excitation, which can happen only
from the uppermost
filled LL,  $ N = 2$. The ML line at $\hbar \omega_{0} + \hbar \omega_{B}$ contributes
to the ${\rm{SU}_{1}}$ satellite if it is accompanied by a $2 \hbar \omega_{B}$ MP excitation, which
can occur from LLs $ N = 2$ and $N=1$. Similarly, the line at
$\hbar \omega_{0} + 2 \hbar \omega_{B}$ contributes to the ${\rm{SU}_{1}}$ satellite
when  a $3 \hbar \omega_{B}$ MP is excited from one of the occupied LLs
$N = 2$,  $N=1$, or $N=0$.

To proceed further, we assume that the hole wave function $\Psi_{h}(\vec{r})$ is
spherically symmetric. Then only zero-angular momentum states contribute
to the optical transitions, $M_{Nm}=M_{N0}  \delta_{m0}$.  The usual selection rule
for interband magneto-optical transitions $N_{electron} = N_{hole}$,
which is valid for a nonlocalized hole, is lifted. For a localized hole,
on the contrary, all LLs can be optically active. Indeed, if the hole is localized on a scale
smaller than $l_{B}$, the optical matrix element $M_{N0}$, according to its definition above,
is given by

\begin{eqnarray}
M_{N0} = P_{vc}\;\int d\vec{r} \;\Psi_{N0}^{\star}(\vec{r})\;\Psi_{h}(|\vec{r}|)
= P_{vc}\;\Psi_{N0}^{\star}(0)\;
\int d\vec{r} \;\Psi_{h}(|\vec{r}|),
\label{Mf}
\end{eqnarray}
where $\Psi_{Nm}(\vec{r})$ is the electron wave function. Thus, the optical matrix elements
entering Eq. (\ref{SAM}) are defined by
the electron wave function at the origin,  $\Psi_{N0}(0) = 1/\sqrt{2 \pi} l_{B}$,  and the hole
localization length $a \equiv \int d\vec{r} \;\Psi_{h}(|\vec{r}|)$, giving
$M_{N0} = P_{vc}\;a/\sqrt{2 \pi} l_{B} \equiv M $. As a result,
the optical transitions from all the populated LLs to the hole level $E_{0}$ are allowed.
We assume also that the electron-hole interaction occurs via a screened potential[17].
In what follows we choose the hole potential to be $V(|\vec{r}|) = V_{0}\;\exp{\{-r^2/2 L^2\}}$
with the potential amplitude $V_{0}$,  and the potential range $L$. Note that
the amplitude $V_{0}$ in this case is defined by two parameters, the hole localization length
 $a$ and the potential range $L$. Depending on the potential amplitude $V_{0}$,
the localization length $a$ can be larger or smaller than the potential range $L$. In the
first case the hole potential is short ranged, while in the second case it can be long ranged.
Since the interaction potential $V(|\vec{r}|)$ is spherically symmetric, the electron scattering
takes place only between
Landau states with the same angular momentum, $V_{Nm, N'm'}=V_{Nm,N'm} \delta_{mm'}$.

Under the assumptions above we obtain for the SU spectrum

\begin{eqnarray}
{\cal {A}}_{k} =  |M|^2 \; \sum_{N,N'} \;
\sum_{m} \frac{|V_{(N+N'+k) m, N'm}|^{2}}{(\varepsilon_{N+N'+k} -
\varepsilon_{N'})^{2}} \;  F_{NN'k} ,\nonumber\\
{\cal {M}}_{k} =  |M|^2 \; \sum_{N,N'} \;\frac{V_{N0, (N+N'+k)0} V_{(N+N'+k)0, N'0}}
{(\varepsilon_{N+N'+k} - \varepsilon_{N})
(\varepsilon_{N+N'+k} - \varepsilon_{N'})}\; F_{NN'k}.
\label{SUs}
\end{eqnarray}
All angular momentum channels contribute to the shakeup term ${\cal {A}}_{k}$,  as all states are
shaken up by the disappearance of the hole in the optical transition. On the other hand,
only the zero-angular momentum channel contributes to the exciton term ${\cal {M}}_{k}$,
since only this
channel is involved in the optical process. Thus, in the general case, the ${\rm {SU}}$ spectrum
is mostly due to the Fermi sea shakeup effect.

\section{Short range hole potential}
\label{sec: srp}

When the hole potential $V(|\vec{r}|)$ is short range compared to $l_{B}$, $L \ll l_{B}$,
nonzero-angular momentum shakeup processes are strongly suppressed. Indeed, at  $L \ll l_{B}$,
using  $\Psi_{Nm}(0) = \delta_{m0}/\sqrt{2 \pi} l_{B}$ one obtains

\begin{eqnarray}
V_{Nm, N'm} = \int d \vec{r} \; \Psi_{Nm}(\vec{r})\; V(|\vec{r}|)\;
\Psi_{N'm}^{\star}(\vec{r})\nonumber\\
= \Psi_{Nm}(0)\;\Psi_{N'm}^{\star}(0)\;
\int d \vec{r} \;  V(|\vec{r}|)
=  V_{0} \; \frac{L^2}{2 \pi l_{B}^2}\; \delta_{m0}.
\label{V1}
\end{eqnarray}
As a result, only zero-angular momentum term survives in Eq. (\ref{SUs}),  and
the intensity of the $k$th satellite reduces to the following simple form:

\begin{eqnarray}
I_{SU_{k}} = I_{ML}^{0} \; \alpha^{2} \;\sum_{N,N'} \Bigl\{\frac{1}{(N + k)^{2}}
- \;\frac{1}{(N + k) (N' + k)}\Bigr\}\;F_{NN'k},
\label{SUl}
\end{eqnarray}
where $I_{ML}^{0} = {\cal{N}} |M|^2$ and the parameter
$\alpha = (V_{0}/2 \pi \hbar \omega_{B})\;(L/l_{B})^{2} = V_{0} \; m_{e} L^2/2 \pi \hbar^2$
defines the interaction strength and is magnetic field independent.

Thus, for a short range hole potential, the shakeup contribution to the ${\rm {SU}_{k}}$ satellite
is strongly suppressed by the excitonic contribution. One can see from Eq. (\ref{SUl})
that only the processes in which optical transitions and  MP excitations occur from different LLs ,
$N \neq N'$, now survive. In this case, if the three lowest LLs are populated, there is no
contribution
to the ${\rm{SU}_{1}}$ satellite from the ML line at $\hbar \omega_{0} + \hbar \omega_{B}$
when a $2 \hbar \omega_{B}$ MP is excited from the  $N=1$ LL (see Fig.~\ref{fig.1}).
In addition, there is no contribution
from the ML line at $\hbar \omega_{0} + 2 \hbar \omega_{B}$ when a $3 \hbar \omega_{B}$ MP
is excited from the $N = 2$ LL.
The most important consequence of the result above is that when only the lowest LL is
occupied, i.e.,  the Landau numbers are $N=N'=0$ in Eq. (\ref{SUl}),
the shakeup spectrum disappears.
This conclusion is, probably, not surprising. Under the conditions above, the problem reduces
effectively to the one-electron problem, since there is only one electron in the state
$|N=0,m= 0\rangle$ which is allowed for both optical and scattering transitions.
As was noted above, a strong suppression of the SU spectrum in the case in which electrons
occupy only the lowest LL,  was observed in Ref. \cite{BJ} for GaAs/AlGaAs quantum wells.

When more than one LL is occupied, the intensity of the $k$th satellite decreases with its
number $k$. When this number is larger than the number of the filled LLs $N_{\mu} = \nu$,
$k \gg \nu$, the ${\rm{SU}_{k}}$ peak intensity decreases as $k^{-4}$,
\begin{eqnarray}
I_{SU_{k}} =   I_{ML}^{0} \; \alpha^{2}  \;
\frac{(\nu - 1) \; \nu^{2}\;(\nu + 1)}{12\;k^{4}}.
\label{lk}
\end{eqnarray}
In Eq. (\ref{lk}) the filling factor $\nu$ is taken for the spinless case.

As it follows from Eq. (\ref{SUl}), the intensity of the each ${\rm SU_{k}}$ satellite
increases linearly with the magnetic field $B$ due to the optical matrix
element $|M|^{2} \sim B$. There is also an
implicit dependence on $B$  due to the occupation numbers entering  Eq. (\ref{SUl})
via the function $F_{NN'k}$, Eq. (\ref{SAM1}). When the magnetic field $B$ increases, the
upper LL depopulates and its contribution to the ${\rm {SU}}$ spectrum changes.
Therefore, the linear increase in the shakeup intensity can be violated for some
regions of $B$, and a nonmonotonic dependence on the magnetic field can be expected.

As an illustration, the intensities for the first three satellites,  ${\rm{SU}_{1}}$,
${\rm{SU}_{2}}$, and ${\rm{SU}_{3}}$, as a function of the magnetic field $B$ are shown
in Fig.~\ref{fig.2}. The magnetic field $B$ increases from the initial magnitude of
 $B = B_{4/5}$ which corresponds to the location of the Fermi energy $E_{F}$ halfway
between LLs $N=4$ and $N=5$. The occupation numbers entering Eqs. (\ref{SAM1}) and (\ref{SUl})
are modeled by the function
$n(\varepsilon_{N}) \rightarrow \frac{1}{2}\; [1 + \Phi((E_{F} - \varepsilon_{N})/\gamma)]$,
where $\Phi(x)$ is the probability integral \cite {GR},
and $\gamma \ll \hbar \omega_{B}$ is the small phenomenological width of the LL.
The curves in Fig.~\ref{fig.2} are calculated using the parameters
$\gamma / \hbar \omega_{B_{4/5}}= 0.2$ and $\alpha = 0.1$.
It can be seen from Fig.~\ref{fig.2} that the shakeup intensities show a pronounced nonmonotonic
dependence on the magnetic field $B$. The changes are periodic with $B$ and the intensity extrema
appear when the uppermost LL crosses the Fermi level. The oscillations are more pronounced for the
first ${\rm SU_{1}}$ satellite. As was noted above,  periodic changes of the shakeup
intensities as a function of the magnetic field are observed in Ref.\cite{Me} at low
temperatures in a modulation-doped quantum well of InGaAs.

\section{Influence of the potential range}
\label{sec: prif}

It is evident that the condition above for the hole potential to be short range, $L \ll l_{B}$,
can be violated at high $B$ due to the decrease in magnetic length,  $l_{B} \sim B^{-1/2}$.
Thus, the results above can be not valid, for example, when the SU spectrum
is measured at high $B$.
We will study here the influence of the potential range on the SU spectrum and consider first
the case when only the lowest LL is occupied, in which case the potential range is crucial.
Then, calculating the scattering matrix elements and performing the summation over the
angular momentum $m$ in Eq. (\ref{SUs}), with the model potential above, one obtains

\begin{eqnarray}
I_{SU_{k}} =  I_{ML}^{0} \; \frac{V_{0}^2}{(k \hbar \omega_{B})^2}\;
W(L/l_{B}) \; n^{2}(\varepsilon_{0})  .
\label{r1}
\end{eqnarray}
where the function

\begin{eqnarray}
W(x) = x^{4}\;\Bigl[\frac{1}{(2 x^2 + 1)^2} - \frac{1}{(x^2 + 1)^{4}} \Bigr], \qquad
W(x) = \left\{
\begin{array}{cr}
2 x^{8}, & x \ll 1, \\
\frac{1}{4} (1 - 1/x^2), &  x \gg 1.
\end{array}
\right.
\label{W}
\end{eqnarray}

For a short range potential when $L \ll l_{B} $, the function  $W(L/l_{B})$
is very small, see Eq. (\ref{W}), and therefore, the SU spectrum is strongly
suppressed, in agreement with the results above. This is not a case, however, when the potential
is long range and  $L \geq l_{B}$. Then the ${\rm SU_{k}}$ satellite intensity can be not
small. In this case it is determined mostly by the interaction
strength $V_{0}/\hbar \omega_{B}$, and its number $k$.

When the magnetic field $B$ decreases,  the  ${\rm SU_{k}}$ intensity
taken in units of the ML intensity $I_{ML}^{0} n(\varepsilon_{0})$, decreases  as $B^{-2}$
due to the cyclotron energy. The function $W(L/l_{B})$ decreases with increasing $B$, and also
there is an additional decrease $\sim B^{-1}$ due to the occupation numbers $n(\varepsilon_{0})$.
Introducing a phenomenological width for the $N=0$ LL in the same manner as in Fig.~\ref{fig.2},
the occupation numbers $n(\varepsilon_{0})$ are replaced by the filling factor $\nu \sim 1/B$.
Note that the filling factor of the $N=0$ LL tends to zero only at $B \rightarrow \infty$,
as the lowest LL cannot be empty for a given electron density.
Fig.~\ref{fig.3} illustrates the magnetic field dependence of the intensity of the
 ${\rm SU_{1}}$ satellite, Eq. (\ref{r1}), at different potential range parameters
$L/l_{B}$.
The magnetic field  $B$ increases from the initial magnitude of $B=B_{0/1}$ which corresponds
to the location of the Fermi level halfway between the $N=0$ and $N=1$ LLs.
The curves in Fig.~\ref{fig.3} are calculated using the following parameters:
$V/\hbar \omega_{B_{0/1}} = 0.3 $,  and $L/l_{B_{0/1}} = 0.6, 1, 1.4$.
From Fig.~\ref{fig.3} one can conclude the following. When only one LL is occupied,
the ${\rm SU_{1}}$ intensity smoothly decreases with increasing $B$. At high magnetic fields
it is almost independent of the potential range $L$ and is very small. At
low magnetic fields $B \simeq B_{0/1}$, the longer the potential range, the larger the
intensity. In addition,  the decay of the intensity is more pronounced at larger $L$, and
is faster than $B^{-2}$.

If several LLs are filled,  the shakeup intensities are given by
expressions similar to Eq. (\ref{r1}). However, the dependence on the potential range
parameter $L/l_{B}$ is now more complicated, because of the increasing number of
possible shakeup processes. In this case the satellite intensities are proportional to
$(L/l_{B})^2$ at $L \ll l_{B}$, see Eq. (\ref{SUl}), and are almost independent of  $L/l_{B}$ at
$L \gg l_{B}$. Therefore, for a short range potential,  the satellite intensity
Eq. (\ref{SUl}) increases linearly with increasing $B$ due to the optical matrix elements.
When the uppermost LL depopulates, its contribution to the satellite intensity
decreases with increasing $B$ and overbalances the linear increase. As a result, a reduction of the
satellite intensity occurs when ever the uppermost LL crosses  the Fermi energy. On
the contrary, when the potential is long range, the satellite intensity decreases with
increasing $B$ as $B^{-1}$,  due to the optical matrix elements and the cyclotron energy, and
it decreases also with $B$
due to the occupation numbers. Thus, one can expect that the oscillating dependence of the shakeup
intensities on the magnetic field  $B$, shown in Fig.~\ref{fig.2}, will be much less pronounced
for the case of a long range potential. This is illustrated in  Fig.~\ref{fig.4}.

In Fig.~\ref{fig.4} the ${\rm SU_{1}}$ satellite intensity is shown as a function of
 the magnetic field $B$,  for different potential ranges. The Fermi
energy is halfway between the $N=1$ and $N=2$ LLs at the magnetic field $B=B_{1/2}$.
The curves in Fig.~\ref{fig.4} are calculated
using the parameter $\gamma/\hbar \omega_{B_{1/2}} = 0.2$,
$V/\hbar \omega_{B_{1/2}} = 0.3$, and  $L/l_{B_{1/2}} = 1, 1.4, 3.2$
(curves 1, 2, and 3,respectively). The occupation numbers $n(\varepsilon_{1})$ and $n(\varepsilon_{0})$
are modeled in the same manner as in Fig.~\ref{fig.2} and Fig.~\ref{fig.3}.
From Fig.~\ref{fig.4} it can be seen, that the shorter the potential range, the more
pronounced the nonmonotonicity of the ${\rm SU_{1}}$ intensity as a function of $B$,
as expected.
Thus, when the potential range $L \simeq l_{B}$, the ${\rm SU_{1}}$ intensity has a wide peak
at those magnetic fields in which the uppermost LL depopulates, see curves 1, and 2.
However, at larger $L \simeq 3 l_{B}$ it decreases almost smoothly  with increasing $B$, as
is shown in Fig.~\ref{fig.4} by curve 3.

\section{Discussion}
\label{sec: dis}

The results above are obtained for a hole wave function and a hole potential that are spherically
symmetric. They are valid also for a nonsymmetric case,  when the hole potential is short range.
 Indeed, at $L \ll l_{B}$, the scattering matrix elements
are defined only by the properties of the electron wave function at $\vec{r}=0$, but not by
the potential symmetry. It refers also to the optical matrix elements as the localization
length is assumed to be small, $a \ll l_{B}$. In the opposite case, that of a long range
potential,  there are, however, changes. Unlike the previous case of a symmetric potential, now
electron scattering occurs also between different angular momentum states. Therefore,
 additional shakeup channels appear and the results must change quantitatively.
However, the shakeup processes in this case  will dominate, as before. The reason is the same
as above, namely, all the angular momentum states are shaken up, but the excitonic contribution
involves optically active states with only zero angular momentum.

The other restriction is  that the hole localization length is much smaller
than the magnetic length, $a \ll l_{B}$. When this condition is violated, the optical
matrix elements $M_{Nm}$ entering Eq. (\ref{SAM}) are changed. They depend now on the electron
quantum number $N$ and the hole localization length $a$. Choosing the hole wave function
to be  $\Psi(|\vec{r}|) = (2 a)^{-1}\; \exp{\{-r^2/4 a^2\}}$, one obtains

\begin{eqnarray}
M_{Nm} = P_{vc}\; \frac{l_{B}}{\sqrt{2 \pi} a}\; \Bigl( \frac{l_{B}^2}{a^2} - 1 \Bigr)^{N} \;
\Bigl( \frac{l_{B}^2}{a^2} + 1 \Big)^{-N-1} \; \delta_{m0}.
\label{Mn}
\end{eqnarray}
In addition, the scattering matrix elements Eq. (\ref{V1}) have to be substituted by

\begin{eqnarray}
V_{N m S m_{s}} = \int \; d \vec{r_{1}} d \vec{r_{2}} \;
\Psi_{N m}(\vec{r_{1}}) \Psi_{S m_{s}}^{\star}(\vec{r_{1}}) \; V(\vec{r_{1}} - \vec{r_{2}})\;
|\Psi_{h}(\vec{r_{2}})|^{2}.
\label{Vn}
\end{eqnarray}
The matrix elements (\ref{Vn}) describe the Auger-like electron-hole scattering processes.
When $a$ is not small compared to $l_{B}$, the scattering
matrix elements (\ref{Vn}) are finite for all angular momentum, even for a short range hole potential.
Thus, the shakeup processes are now  always dominant and
the SU spectrum exists in this case also at filling factors $\nu \leq 1$. To demonstrate this
consider, for example,  the case of  $a \simeq l_{B}$. In this case the optical transitions
from the lowest LL  $N=0$, are the most intensive,
$M_{Nm} \simeq M_{0}\;\delta_{N0}\;\delta_{m0}$, where $M_{0} = P_{vc}/2 \sqrt{2 \pi}$.
There is only one line in the main ML spectrum
with  energy  $\hbar \omega_{0}$ and, therefore,  the number of contributions
to the SU spectrum reduces. In  Fig.~\ref{fig.1}, for example,  only one contribution
to the ${\rm SU_{1}}$ survives, when the ML line at $\hbar \omega_{0}$ is accompanied by
an $\hbar \omega_{B}$ MP excitation from the  $N=2$ LL.
The intensity of the ${\rm SU_{k}}$ satellite at $a \simeq l_{B}$ is given by

\begin{eqnarray}
I_{SU_{k}} =  {\cal {N}} \; |M_{0}|^2 \; \frac{1}{(k \hbar \omega_{B})^2}\;\Bigl\{ \sum_{N,m}
|V_{Nm (N+k)m}|^2 \; F_{0Nk} - |V_{00 k0}|^2 \; F_{00k} \Big\},
\label{Sdh}
\end{eqnarray}
where $N$ numbers the occupied LLs. The shakeup contribution is dominant and, moreover,
the excitonic term does not contribute to
each ${\rm SU_{k}}$ satellite. Indeed, the function $F_{00k}$ differs from zero only
if the $k$th LL is empty, see Eq. (\ref{SAM1}). Therefore, the excitonic term contributes
only to the satellites,  with a number $k$ equal or larger than the number of occupied LLs.
For example, when the three lowest LLs $N=0$, $N=1$, and $N=2$
are occupied, as in Fig.~\ref{fig.1}, this term do not contribute to the
first ${\rm SU_{1}}$ and second  ${\rm SU_{2}}$ satellites, only to the
third and higher satellites. Thus, a strong hole localization is not a necessary requist for
the SU spectrum to be realized. It is not a case, however, for the FES effect. As it follows
from Eq. (\ref{PE}), in the case of $a \simeq l_{B}$ the many-body corrections to the main
spectrum vanish, as only the lowest Landau state is optically active and the optical transitions
from the higher LL $N\neq0$ are much less intensive, $M_{N0} \simeq 0$.
The result above for the SU spectrum is in contradiction to the statement in the early experimental
works (see, e.g., Refs.\cite{N},\cite{NS}) that both the FES effect and SU spectrum are very
sensitive to a degree of the hole localization. Our conclusions are, however, in agreement with
the results from Refs.\cite{US} and \cite{Me2}. The numerical calculations in Ref.\cite{US} (see
Fig.1(b) in Ref.\cite{US}) and recent experimental observations in Ref.\cite{Me2} show that the
SU spectrum, contrary to the FES effect, do not depend stronly on the hole localization.

The effect of the electron spin can be accounted for if we assume that the spin of the conduction
electron is conserved in the scattering event with the localized hole. In fact this approximation
is implicit in the statement above that only ${\it one}$ localized level is involved.
Under such a restriction the exchange scattering processes in which the conduction electron and
the localized hole would both reverse their spin, are ignored. In this case the only change is that
the emission intensities Eqs. (\ref{PE}) and (\ref{SU}) are multiplied by a factor 2 as there are
twice as many electrons involved.

The results above can not be extended to the case of delocalized holes. In this case, as well as
in the case of a many state localized hole,
the Nozi\` eres - De Dominicis formalism, i.e.,  the ``one-body'' representation for the
two-particle Green's function, is not valid and to obtain the emission spectrum one has to calculate
the two-particle Green's function.

The case of delocalized holes is studied in Ref.\cite{BJ},  where a  strong suppression of the
${\rm SU_{1}}$ satellite at $\nu \leq 1$ was observed for GaAs/AlGaAs quantum wells,
in which the holes are assumed to be unlocalized. The amplitude for the shakeup
optical transition was calculated up to the second order in the perturbation, namely, up to
the first order in the coupling with electromagnetic field, and up to the first order in
the electron-hole or electron-electron interaction. The states of the electron system
are given by Slater determinants. It was shown that when an electron from the lowest LL recombines
with a valence hole, the contributions to the shakeup transition amplitude due to the
electron-hole and electron-electron scattering processes cancel each other. Such a mechanism
for the SU emission quenching differs from the one presented above. In our model the electrons
in the conduction band are assumed to be noninteracting and the SU spectrum disappears as there are
no shakeup channels when the hole potential is short range.

\section{Conclusion}
\label{sec: con}

The emission spectrum due to transitions between a localized level and a 2DEG in a strong
magnetic field is calculated analytically. Calculations are performed with a
Nozi\` eres-De Dominicis-like Hamiltonian where the interaction part is treated as a perturbation,
since the hole potential is assumed to be small compared to the cyclotron energy,
$V_{0} \ll\hbar \omega_{B}$. The shakeup spectrum appears only when the emission is
calculated up to the second-order in this perturbation and is a result of a balance
between the orthogonality catastrophe and the excitonic effect contributions. In the general case the
shakeup effect is dominant as all the angular momentum states are shaken up, while
the excitonic contribution involves the optically active states with zero angular momentum only. For
the special case of a short range hole potential, however, the excitonic effect competes with
the shakeup effect as nonzero-angular momentum shakeup processes are
strongly suppressed. If electrons occupy only the lowest LL, i.e.,  the filling factor
$\nu \leq 1$, there are no channels to shake up the Fermi sea in the optical transition
and the SU spectrum disappears. When electrons occupy two or more LLs, the satellite intensities
oscillate as a function of the magnetic field. For the opposite case, that of a long range hole
potential,
the shakeup spectrum is determined mostly by the Fermi sea shakeup processes,  and
the excitonic contribution is much less important. In this case the satellite intensities
decrease smoothly when the magnetic field increases and there is no suppression
of the SU spectrum when electrons occupy the lowest LL only.
A strong hole localization, $a \ll l_{B}$, is not a crucial condition for the SU spectrum
to be observed. If the hole localization length is not small compared to $l_{B}$,  the SU spectrum
still exists. Only the number of contributions to the SU spectrum
reduces and the shakeup processes are always dominant, as compared with the case of a strong
 hole localization.

\section{Acknowledgments}
\label{sec: ac}

This work was supported by the MINERVA Foundation. One of us (E. T.) wishes to thank the
Department of Condensed Matter Physics of the Weizmann Institute of Science for its
support and hospitality. One of the authors (Y. L.) thanks I. Bar-Joseph for discussions
that stimulated this work.

\begin{figure}
\caption{A schematical illustration of different contributions to the first
 $SU_{1}$  shakeup satellite. The solid lines show the occupied LLs which are also opticaly active.
The dashed lines show the empty LLs. The solid and the dashed arrows show the optical transitions and
the magnetoplasmon creations, respectively.}
\label{fig.1}
\end{figure}
\begin{figure}
\caption{The intensities of shakeup satellites as a function of the magnetic field $B$,
for a short range potential.  The intensities are given in units of $I_{ML}^{0}$ at $B=B_{4/5}$.}
\label{fig.2}
\end{figure}
\begin{figure}
\caption{The intensity of the first $SU_{1}$ shakeup satellite as a function of the magnetic
field $B$ for different localization parameters. The parameter
$L/l_{B_{0/1}} = 1.4, 1, 0.6$ for curves 1, 2, and 3, respectively.
At $B=B_{0/1}$ electrons occupy the lowest LL $N=0$. The intensity is given in units of
$I_{ML_{0}}^{0} n(\varepsilon_{0})$.}
\label{fig.3}
\end{figure}
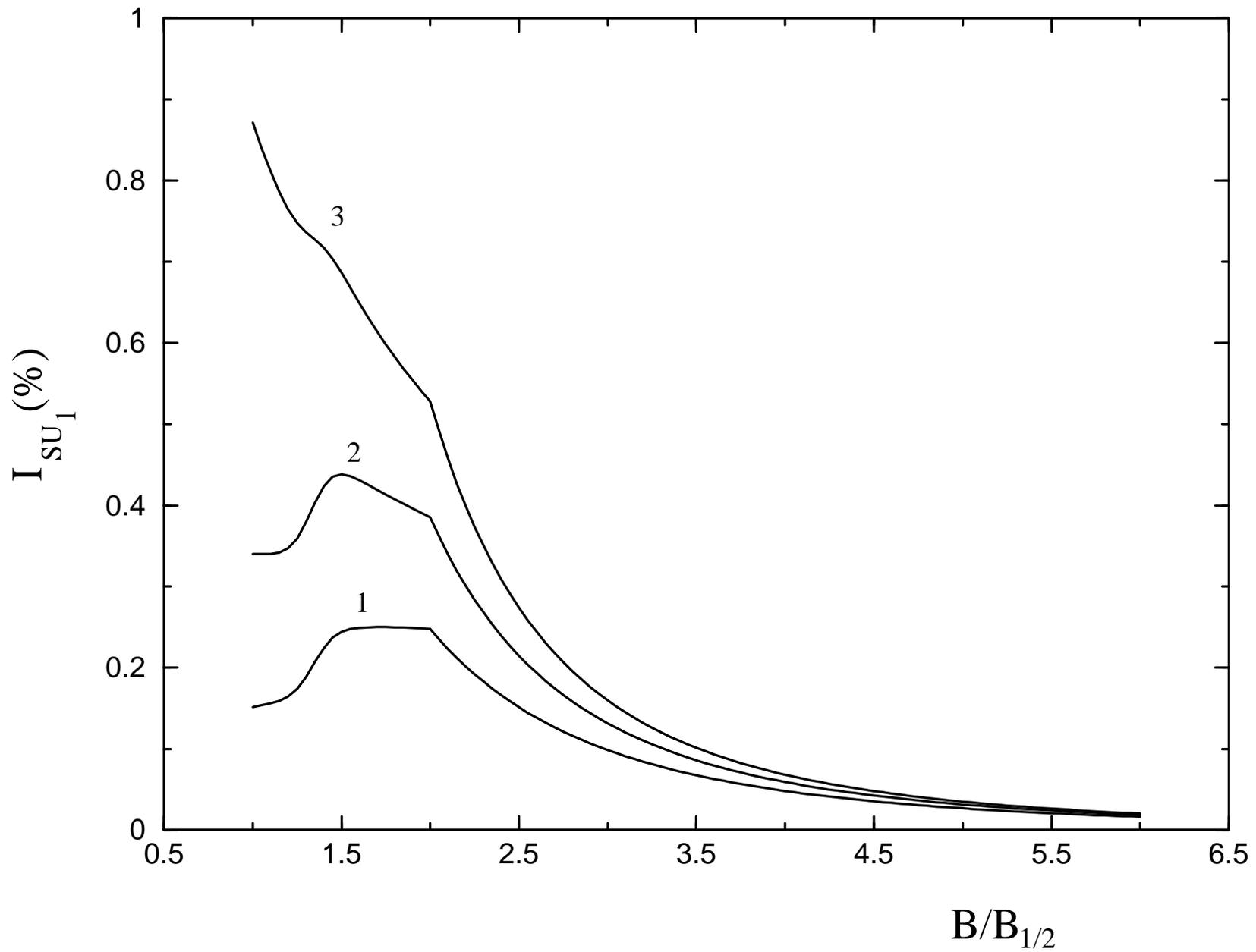
\begin{figure}
\caption{The intensity of the first $SU_{1}$ shakeup satellite as a function of the magnetic
field $B$  for different localization parameters.  The parameter
$L/l_{B_{0/1}} = 1, 1.4, 3.2$ for curves 1, 2, and 3, respectively.
At  $B=B_{1/2}$ electrons occupy the two lowest LLs $N=0$, and $N=1$.
The shakeup intensity is given in units of $I_{ML}^{0}$ at $B=B_{1/2}$.}
\label{fig.4}
\end{figure}

\end{document}